\documentclass[amsfonts, amssymb, amsmath, showkeys, nofootinbib, preprint,floatfix, prl, superscriptaddress]{revtex4-2}

\usepackage[colorinlistoftodos, color=green!40,prependcaption]{todonotes}
\usepackage{graphicx}
\usepackage{dcolumn}
\usepackage{bm}
\usepackage{physics}
\usepackage{float}
\usepackage{xcolor}
\usepackage{amsmath}
\usepackage{amssymb}
\usepackage{units}
\usepackage[english]{babel}
\usepackage{placeins}
\usepackage{lineno}

\usepackage{lipsum}

\begin{document}

\title{Electron-magnon coupling \\at the interface of a ``twin-twisted'' antiferromagnet}

\author{Yue Sun}
\thanks{Y.S. and F.M. contributed equally to this work.}
\affiliation {Department of Physics, University of California, Berkeley, California 94720, USA}
\affiliation {Materials Science Division, Lawrence Berkeley National Laboratory, Berkeley, California 94720, USA}

\author{Fanhao Meng}
\thanks{Y.S. and F.M. contributed equally to this work.}
\affiliation {Materials Science Division, Lawrence Berkeley National Laboratory, Berkeley, California 94720, USA}
\affiliation {Department of Materials Science and Engineering, University of California, Berkeley, California 94720, USA}

\author{Sijia Ke}
\affiliation {Department of Materials Science and Engineering, University of California, Berkeley, California 94720, USA}
\affiliation {Chemical Science Division, Lawrence Berkeley National Laboratory, Berkeley, California 94720, USA}

\author{Kun Xu}
\affiliation {Department of Mechanical Engineering, Stanford University, Stanford, CA, USA}

\author{Hongrui Zhang}
\affiliation {Department of Materials Science and Engineering, University of California, Berkeley, California 94720, USA}

\author{Aljoscha Soll}
\affiliation {Department of Inorganic Chemistry, University of Chemistry and Technology Prague, Technická 5, 166 28 Prague 6, Czech Republic}

\author{Zden\v ek Sofer}
\affiliation {Department of Inorganic Chemistry, University of Chemistry and Technology Prague, Technická 5, 166 28 Prague 6, Czech Republic}

\author{Arun Majumdar}
\affiliation {Department of Mechanical Engineering, Stanford University, Stanford, CA, USA}

\author{Ramamoorthy Ramesh}
\affiliation {Department of Physics, University of California, Berkeley, California 94720, USA}
\affiliation {Materials Science Division, Lawrence Berkeley National Laboratory, Berkeley, California 94720, USA}
\affiliation {Department of Materials Science and Engineering, University of California, Berkeley, California 94720, USA}

\author{Jeffrey B. Neaton}
\affiliation {Department of Physics, University of California, Berkeley, California 94720, USA}
\affiliation {Materials Science Division, Lawrence Berkeley National Laboratory, Berkeley, California 94720, USA}
\affiliation {Kavli Energy NanoScience Institute at Berkeley, Berkeley, California 94720, USA}

\author{Jie Yao}
\affiliation {Materials Science Division, Lawrence Berkeley National Laboratory, Berkeley, California 94720, USA}
\affiliation {Department of Materials Science and Engineering, University of California, Berkeley, California 94720, USA}

\author{Joseph Orenstein}
\affiliation {Department of Physics, University of California, Berkeley, California 94720, USA}
\affiliation {Materials Science Division, Lawrence Berkeley National Laboratory, Berkeley, California 94720, USA}

\begin{abstract}
We identify a ``twin-twist'' angle in orthorhombic two-dimensional magnets that maximizes interlayer orbital overlap and enables strong interfacial coupling. Focusing on the van der Waals antiferromagnet CrSBr, we show that this twist angle, near 72$^{\circ}$, aligns diagonal lattice vectors across the layers, enhancing the interlayer hopping that is spin-forbidden in pristine systems and orbital-forbidden in 90$^{\circ}$-twisted samples. The enhanced hopping modifies the electronic structure and activates a novel mechanism for excitation of interfacial magnons.  Using optical probes we discover that excitons on one side of the interface selectively excite magnons localized on the opposite side. We show that this cross-coupling phenomenon can be understood as a consequence of the spin-transfer torque as that arises as electrons tunnel across the twin-twisted interface.  Our findings demonstrate that large-angle twisting in anisotropic 2D materials offers a powerful tool for engineering spin and charge transport through controlled interlayer hybridization, opening new avenues for twisted magnetism and strongly correlated moiré physics.
\end{abstract}

\maketitle

\section{Introduction}
The demonstration of unexpected phases in twisted layers of van der Waals materials such as graphene and transition metal chalcogenides has opened new frontiers in condensed matter physics.   Research has focused for the most part on the electronic \cite{andreiMarvelsMoireMaterials2021} and magnetic \cite{burchMagnetismTwodimensionalVan2018a,makProbingControllingMagnetic2019b} properties of layers whose 2D Bravais lattices are hexagonal or triangular.  However, opportunities for discovery in twisted structures comprised of square or rectangular 2D lattices are, on the whole, less well explored.  

A highly promising candidate is CrSBr, a magnetic van der Waals system that is receiving heightened attention because of its unique optical, electronic, and magnetic properties \cite{ziebelCrSBrAirStableTwoDimensional2024,brennanImportantElementsSpinExciton2024}.  CrSBr is an orthorhombic semiconductor with an optical absorption gap of 1.36 eV. Below a N\'eel temperature of 135 K, CrSBr exhibits layered antiferromagnetic order, in which the magnetization of ferromagnetic bilayers alternates along the layer-stacking direction \cite{GOSER1990129,telfordLayeredAntiferromagnetismInduces2020}.  An easy-plane anisotropy leads to spin orientation within the layer planes.  In addition, the spins experience a weaker anisotropy within the plane, in which the $b$-axis is the preferred direction \cite{lopez-pazDynamicMagneticCrossover2022}. The magnetic field dependence of magnetization indicates that the inter-plane exchange coupling is weak and therefore AFM order is only slightly preferred over FM alignment of neighboring Cr bilayers. 

The transport and optical properties of CrSBr are highly anisotropic as well.  From density functional theory (DFT) calculations, the electronic bandwidths for wavevectors perpendicular to the layer planes are approximately 50 meV \cite{chenTwistassistedAllantiferromagneticTunnel2024}, a factor of about twenty smaller than the valence-band width for wavevectors in the plane. DFT calculations also indicate that the electronic coupling is highly anisotropic within the layer plane as well. In particular, the conduction band is nearly flat throughout the Brillouin zone for wavevectors in the $\Gamma$-$\mathrm{X}$ direction, which corresponds to the orthorhombic $a$-axis, but has a bandwidth of order 1 eV for wavevectors along the $\Gamma$-$\mathrm{Y}$ direction \cite{wilsonInterlayerElectronicCoupling2021a}.  The quasi-one-dimensional character of the conduction band is supported by optical absorption, emission, and photoconductivity measurements \cite{wuQuasi1DElectronicTransport2022,kleinBulkVanWaals2023}.  For example, the ratio of photoluminescence intensity generated by light with polarization parallel and perpendicular to the $b$-axis is approximately $10^2$ \cite{wuQuasi1DElectronicTransport2022,kleinBulkVanWaals2023}.  A similar ratio is obtained for the photocurrent, which reflects the relative efficiency of free-carrier generation \cite{wuQuasi1DElectronicTransport2022}.

The unusual combination of electronic and magnetic anisotropy motivates the search for new phenomena arising in twisted bilayers of CrSBr.  Indeed, fascinating effects of twisting on the magnetic structure at the interface \cite{boix-constantMultistepMagnetizationSwitching2024} and the tunneling magnetoresistance \cite{chenTwistassistedAllantiferromagneticTunnel2024} have already been reported.  In this study we utilize the strong anisotropy in the response to linear polarized light to probe new phenomena arising at the interface between twisted flakes. Using photoluminescence (PL) spectroscopy and time-resolved reflectance we have identified a special twisting angle proximate to which unexpected photophysical phenomena and forms of electron-spin coupling emerge. This angle, $\theta_t$, is given by,
\begin{equation}
\theta_t = 2\tan^{-1}\frac{a}{b},
\end{equation}
where $a$ and $b$ are the lattice parameters of the 2D rectangular unit cell.  At $\theta_t$, the long diagonals of the twisted layers are parallel and the fraction of coincident lattice sites is maximal. As this is the condition that determines the lowest energy direction for a grain boundary, we refer to $\theta_t$ as the ``twin-twisted" angle.  

\section{Experimental results}
\subsection{Photoluminescence}
\begin{figure}[tb]
\centering
\includegraphics[width=1\columnwidth]{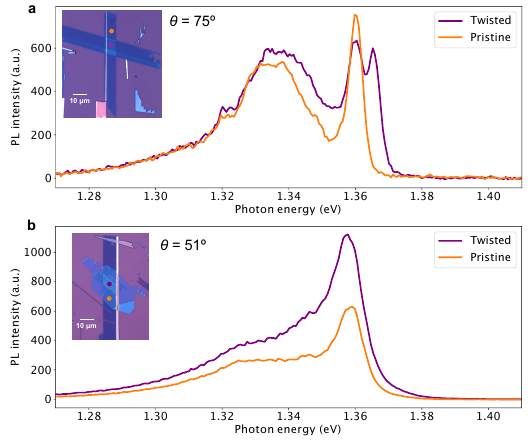}
\caption{Comparison of PL spectra between overlapped regions (twisted) and individual layers (pristine) for (a) 75$^{\circ}$ and (b) 51$^{\circ}$ twist angles. Insets show optical images of both twisted samples, with purple and orange circles marking the measurement locations for the twisted and pristine regions, respectively.}
\label{fig:Fig1_PL_data}
\end{figure}

We now describe characterization of electronic and magnetic excitations in twisted CrSBr using PL spectroscopy and transient reflectance. All twisted samples mentioned in this work are twisted double tetralayer CrSBr (4+4). In the following, we refer to a tetralayer as a ``layer'' for simplicity. By focusing the laser to a spot size smaller than the dimension of the individual crystals, we are able to interrogate each layer individually, as well as the region in which they overlap, by shifting the laser focus between the two positions indicated in the micrographs inset to Figs.~\ref{fig:Fig1_PL_data}. Fig.~\ref{fig:Fig1_PL_data}(a) compares PL spectra from the single-tetralayer (``pristine'') and overlapped regions of a sample with twist angle 75$^{\circ}$, which is close to the twin-twisted angle of 72.65$^{\circ}$. The PL spectrum of the pristine region (red color) shows a broad peak centered at 1.34 eV and a sharper one at 1.36 eV, consistent with previous measurements \cite{marques-morosInterplayOpticalEmission2023}. Upon moving the focus to the overlapped region, we observe an additional narrow peak that is shifted to higher photon energy by 5 meV.  We do not observe this additional peak at the other twist angles we have investigated, which includes a series of small angle twists below 10$^{\circ}$, as well as twists of 51$^{\circ}$, and 90$^{\circ}$.  An example of the data from these measurements is shown as Fig.~\ref{fig:Fig1_PL_data}(b), which compares the PL from the individual layers and the overlapped region for a 51$^{\circ}$ twist [for PL spectra at other twist angles, see Supplementary Information Sec. 1].

\subsection{Spin wave spectroscopy}
We show below that the new features that emerge in the exciton spectrum of twin-twisted CrSBr are accompanied by changes in the magnon spectra. As demonstrated previously, long-wavelength spin waves in single crystals of CrSBr are excited by light pulses whose photon energy is at, or above, the optical gap \cite{baeExcitoncoupledCoherentMagnons2022b}. The subsequent coherent spin precession induces synchronous modulation of the reflectance ($\Delta R$) when probed near the onset of absorption.   A modulation of reflectance that is linearly proportional to small amplitude precession requires canting the spins by the application of a static magnetic field \cite{wilsonInterlayerElectronicCoupling2021a}. This leads to a useful selection rule, whereby the lower spin wave branch is detected when magnetic fields are applied along the $c$-axis while the upper branch is detected for magnetic fields parallel to the $a$-axis \cite{baeExcitoncoupledCoherentMagnons2022b,sunDipolarSpinWave2024b}. 

\begin{figure*}[tb]
\centering
\includegraphics[width=1\linewidth]{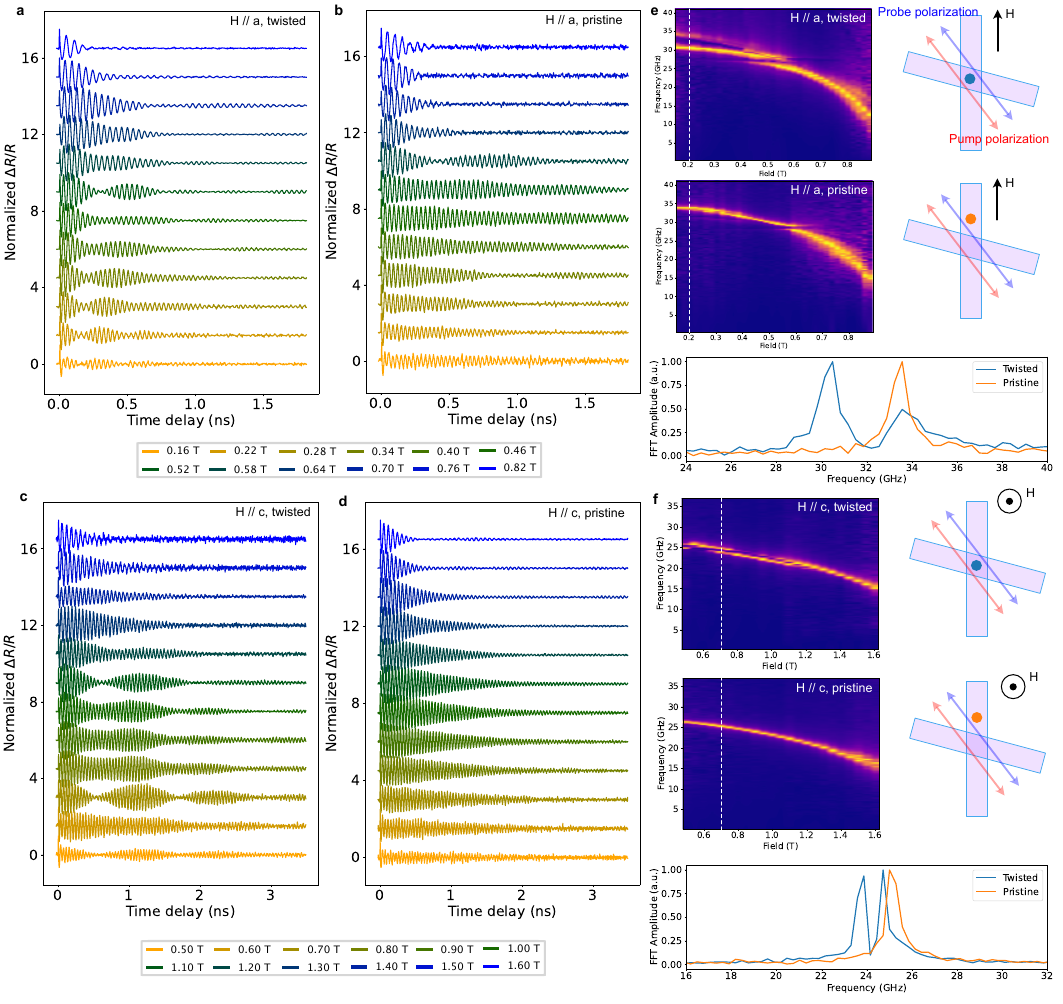}
\caption{(a–d) Comparison of transient reflectance in twisted and pristine regions under different field orientations. Measurements in twisted regions are shown in (a,c), and in pristine regions in (b,d). (a,b) correspond to an in-plane field along the $a$-axis of a composite flake, as indicated by the schematics in the top row (red and blue arrows represent the polarizations of the pump and probe beams, respectively). (c,d) correspond to an out-of-plane field, with schematics shown in the bottom row. (e,f) Field dependence of magnon frequencies for (e) in-plane and (f) out-of-plane fields, obtained via Fourier transform of the data in (a–d). The two linecuts compare twisted and pristine regions under a 0.2 T in-plane field (top row) and a 0.7 T out-of-plane field (bottom row).}
\label{fig:Fig2_Magnon_results}
\end{figure*}

In general, the magnon-related oscillations of the transient reflectance are superposed on a slowly varying component.  In Fig.~\ref{fig:Fig2_Magnon_results} we focus on the oscillatory component of the transient reflectance, $\Delta R_\Omega(t)$, as isolated by a background subtraction procedure described in the Supplementary Information Sec. 2. $\Delta R_\Omega(t)$ observed when pump and probe are in the overlapped region of the twin-twisted sample [Figs.~\ref{fig:Fig2_Magnon_results}(a) and (c)] is compared with measurements that probe the pristine individual layers [Figs.~\ref{fig:Fig2_Magnon_results}(b) and (d)]. The qualitative difference in the two sets of data is quite striking; a clear beating pattern emerges in the transients recorded from the twisted region, suggesting that the magnon peak in the pristine layer is split into two in the overlapped region. The magnetization amplitude in the frequency-field plane, as obtained by Fourier analysis of $\Delta R_\Omega(H,t)$, is compared for the pristine and overlapped regions in Figs.~\ref{fig:Fig2_Magnon_results}(e) and (f), accompanied by a sampling of the spectra at a fixed field of 0.7 T.  We see that magnon modes are split in the overlapped region in both the upper and lower branches of the spin wave spectrum, analogous to splitting of the exciton resonance described previously.

\subsection{Coupling of electronic and magnetic modes}

In pristine CrSBr, the coupling between electrons and magnons is understood to be a consequence of spin-dependent tunneling of electrons between antiferromagnetically oriented layers of spins \cite{wilsonInterlayerElectronicCoupling2021a}. Although tunneling between adjacent layers with opposite spin is forbidden, interlayer tunneling becomes allowed when the spins deviate from antiparallel. The dynamical spin canting associated with coherent magnons generates a nonzero tunneling matrix element that perturbs the electronic levels of each layer. 

\begin{figure}[tb]
\centering
\includegraphics[width=1\columnwidth]{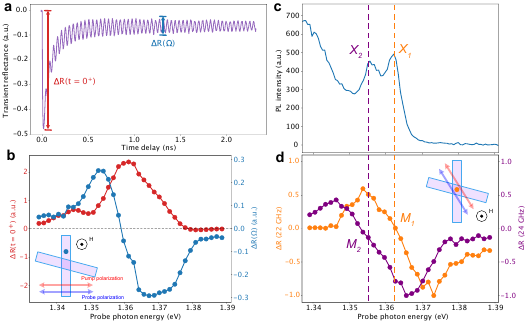}
\caption{(a) Transient reflectance of a pristine sample under a 1.0 T out-of-plane field, with pump and probe beams polarized along the $b$-axis. $\Delta R(t=0^{+})$ represents the initial negative reflectance change, while $\Delta R(\Omega)$ denotes the amplitude of the oscillatory component at frequency $\Omega$. (b) Probe photon energy dependence of $\Delta R(t=0^{+})$ and $\Delta R(\Omega)$ for a pristine sample. (c) PL spectrum measured under a 1.0 T out-of-plane field, with $X_1$ and $X_2$ marking the two PL peaks. (d) Probe photon energy dependence of $\Delta R(\Omega)$ for the two magnon modes, $M_1$ and $M_2$, in a 75$^{\circ}$ twisted sample. The measurement was performed with the pump and probe polarizations aligned along the direction shown in the inset under a 1.0 T out-of-plane field. Dashed lines indicate the coupling between $X_{1,2}$ and $M_{1,2}$.}
\label{fig:Fig3_spin_exciton_coupling}
\end{figure}

This picture of electron-magnon coupling in pristine CrSBr is supported by analysis of the transient reflectance.  We focus on the two features of the transient reflectance highlighted in Fig.~\ref{fig:Fig3_spin_exciton_coupling}(a): the initial negative change, $\Delta R(0)$, and the amplitude of the oscillatory component, $\Delta R_\Omega$. $\Delta R(0)$ is a measure of saturation of an electronic transition resulting from electron-hole pairs generated by the pump. The oscillatory component, $\Delta R_\Omega$, probes the shift of the absorption spectrum that is synchronous with pump-induced coherent magnon precession.

The photon energy dependences of $\Delta R(0)$ and $\Delta R_\Omega$ from a pristine region of the sample, measured with pump and probe beams both polarized along the $b$-axis, are compared in Fig.~\ref{fig:Fig3_spin_exciton_coupling}(b). A peak in the spectrum of $\Delta R(0)$, shown as red circles in Fig.~\ref{fig:Fig3_spin_exciton_coupling}(b), occurs at the same energy, 1.36 eV, as the PL emission. The spectrum of $\Delta R_\Omega$, shown as blue circles, is proportional to the derivative of the $\Delta R(0)$ spectrum, consistent with a shift in the absorption peak that accompanies the spin wave precession.  Taken together, the spectra affirm the view that dynamical spin canting associated with coherent magnons induces a synchronous shift of the exciton energy.

To address the spin-exciton coupling in the overlapped region of 75$^{\circ}$ twisted CrSBr, we label the doubled electronic and magnonic excitations $X_{1,2}$ and $M_{1,2}$, respectively. We then ask whether  spin precession associated with mode $M_{1}$, for example, modulates both excitons, $X_{1}$ and $X_{2}$, or is uniquely coupled to one of the two. To address this question we analyzed $\Delta R(t)$ with pump and probe beams focused at the intersection of the two individual layers. The electric field of both the pump and probe beams are aligned with the common diagonal of the two crystals.  As shown previously in Fig.~\ref{fig:Fig2_Magnon_results}, the Fourier transform of $\Delta R_\Omega(t)$ peaks at 22 GHz and 24 GHz. Fig.~\ref{fig:Fig3_spin_exciton_coupling}(d) compares the amplitude of the 22 GHz and 24 GHz components of $\Delta R_\Omega(t)$ as a function of photon energy. Each component yields a derivative shaped spectrum with zero-crossings that are shifted by 5 meV. Fig.~\ref{fig:Fig3_spin_exciton_coupling}(c) shows that the zero-crossings coincide with the PL peaks. We conclude that each of the magnon modes, $M_1$ and $M_2$, modulates the energy of one, and only one, of the two  excitons, $X_1$ and $X_2$. 

\begin{figure*}[tb]
\centering
\includegraphics[width=1\linewidth]{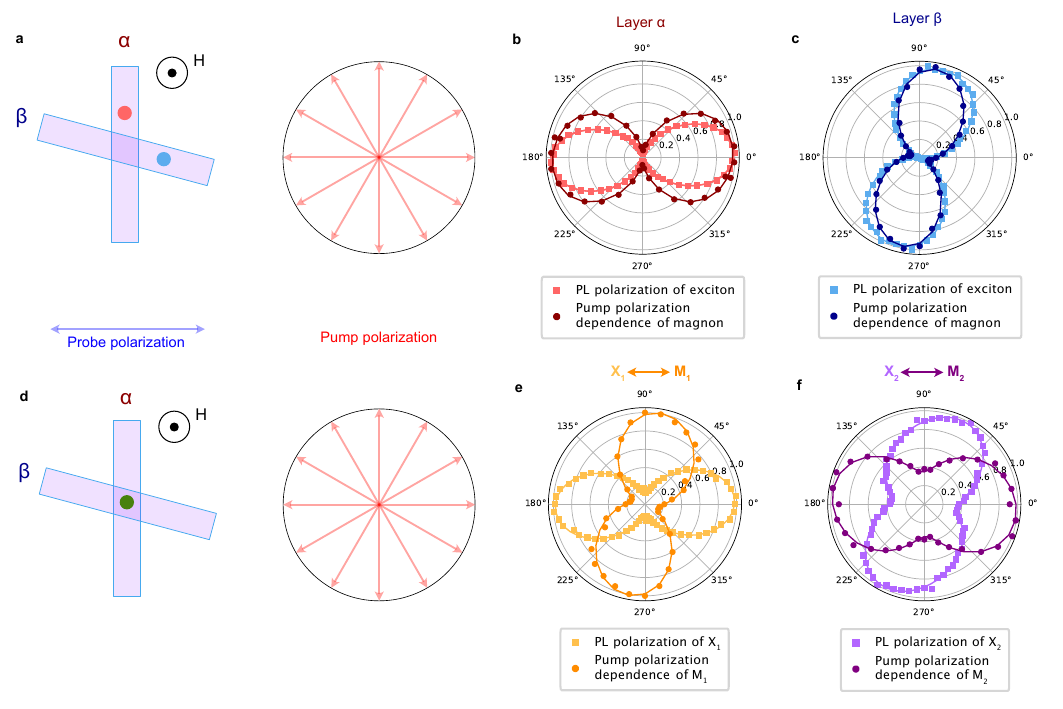}
\caption{(a) Schematic of measurements in pristine regions. $\alpha$ and $\beta$ label the two composite flakes. The transient reflectance measurements of magnon amplitudes are performed with the probe polarization fixed along the $b$-axis of layer $\alpha$ while varying the pump polarization. (b,c) PL polarization of excitons (light-colored squares) and pump polarization dependence of magnons (dark-colored circles) for (b) layer $\alpha$ and (c) layer $\beta$. (d) Schematic of measurements in the twisted region. Pump and probe polarizations are the same as (a). (e,f) PL polarization of excitons (light-colored squares) and pump polarization dependence of magnons (dark-colored circles) for (e) the $X_1M_1$ pair and (f) $X_2M_2$ pair.}
\label{fig:Fig4_cross_coupling}
\end{figure*}

Having established the one-to-one coupling $M_i\leftrightarrow X_i$ we used the strong electronic anisotropy of CrSBr to probe the spatial correlation of the coupled modes. We compare the polarization of the PL with the pump polarization dependence of $\Delta R_\Omega$. The experimental scheme is illustrated in the sketch in Fig.~\ref{fig:Fig4_cross_coupling}(a), where we have labeled the two layers "$\alpha$" and "$\beta$."

We begin with a description of the results obtained with laser focus in the non-overlapped region of layer $\alpha$. The PL amplitude as a function of polarization angle is plotted in Fig.~\ref{fig:Fig4_cross_coupling}(b) as red circles, showing that the transition dipole is oriented along the $b$-axis of layer $\alpha$. In the same plot, we show (as brown circles) $\Delta R(\Omega)$ as a function of the direction of polarization of a pump beam.  Fig.~\ref{fig:Fig4_cross_coupling}(c) presents the same set of measurements performed in layer $\beta$.  In agreement with previous results on single crystals, the amplitudes of both PL and $\Delta R(\Omega)$ are maximized along the $b$-axis of probed layer, which is the direction of its strongest emission and absorption.

In Figs.~\ref{fig:Fig4_cross_coupling}(e) and (f) we present measurements of the polarization dependence in the overlapped region. If the exciton splitting in this region resulted from hybridization across the interface, the transition-dipole would be oriented along the mutual diagonal.  Instead, we find that the transition dipoles extracted from the polar plots align with the $b$-axes of $\alpha$ and $\beta$, indicating that excitons and magnons in the twisted system remain largely localized within individual layers. From our ab initio \textit{GW} and Bethe-Salpeter equation (\textit{GW}-BSE \cite{deslippe2012}) calculations of model twin-twisted bilayer structures, the two lowest excitons are bright, dominated by transitions of non-zero $k$ due to the strong electron-hole interactions, and localized in one or the other constituent layers, respectively. Our calculated transition dipole matrix elements are consistent with the PL measurements of their polarization (see Supplementary Information Sec. 4). 

As excitons are each strongly localized within one layer, we may relabel them as $X_{\alpha,\beta}$ rather than $X_{1,2}$. It would be natural to suppose that the coupling between exciton-magnon pairs would be of the form $X_{\alpha,\beta}\leftrightarrow M_{\alpha,\beta}$, in other words excitons and magnons in the same layer should be coupled more strongly than a cross-layer pair. Consequently, the polarization patterns for the overlapped regions would be expected to resemble those in Figs.\ref{fig:Fig4_cross_coupling}(b) or (c). However, the data shown in Figs.~\ref{fig:Fig4_cross_coupling}(e) and (f) defy this expectation. 

Fig.~\ref{fig:Fig4_cross_coupling}(e) compares the polarization of the $X_\alpha$ PL peak and that of the magnon mode (at 22 GHz) to which it is strongly coupled (as previously demonstrated in Fig.~\ref{fig:Fig3_spin_exciton_coupling}). The transition dipole of $X_\alpha$ is parallel to the $b$ axis of layer $\alpha$. If the magnon to which it couples were co-located in layer $\alpha$, the spin precession amplitude would be largest when the pump beam is polarized in the same direction. Instead, as seen in Fig.~\ref{fig:Fig4_cross_coupling}(e), the generation of the 22 GHz magnon is maximized when the pump polarization angle is oriented parallel to the $b$-axis of layer $\beta$. Fig.~\ref{fig:Fig4_cross_coupling}(f) shows that the same anomaly appears in the coupling of $X_\beta$ to the 24 GHz magnon. Thus we find, surprisingly, that the exciton localized in layer $\alpha$ is coupled most strongly to the magnon localized in layer $\beta$, and \textit{vice versa}, a result we can summarize symbolically as $M_{\alpha,\beta} \leftrightarrow X_{\beta,\alpha}$.


\subsection{Moir\'e superlattice}

The observation of cross-layer coupling, $M_{\alpha,\beta} \leftrightarrow X_{\beta,\alpha}$, indicates that the exciton-magnon interaction in the overlapped region is dominated by processes occurring at the interface between layers. As the phenomena of doubling of modes appears only near the twin-twist angle, we conclude that that the moire pattern created by $\theta_t$ at the interface plays a critical role. As we discuss below, the moire lattice displays several unique features for twist angles close to $\theta_t$. 

\begin{figure}[tb]
\centering
\includegraphics[width=0.9\columnwidth]{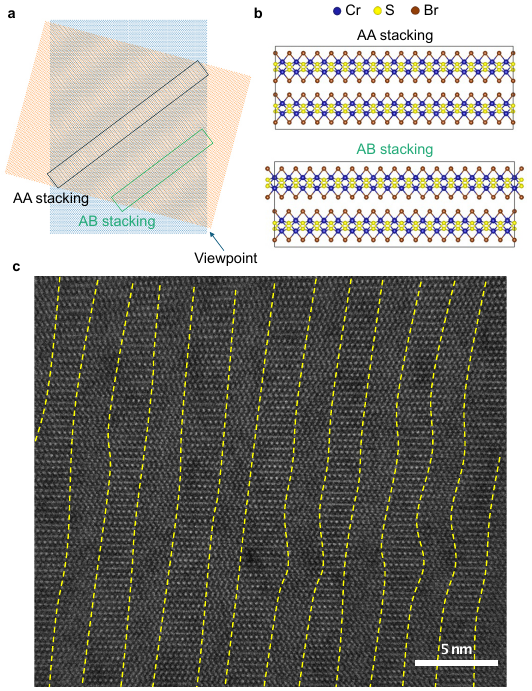}
\caption{(a) Top view of the crystal structure of 75$^{\circ}$-twisted CrSBr. Each dot represents a Cr atom, with the top and bottom layers colored in orange and blue, respectively. (b) Crystal structures of AA and AB stackings viewed along the arrow in (a). (c) 1D moiré pattern observed in a TEM image of 75$^{\circ}$-twisted CrSBr. AA and AB stackings are delineated by yellow dashed lines.}
\label{fig:Fig5_moire_pattern}
\end{figure}

When $\theta=\theta_t$, the diagonals of the unit cells of the two CrSBr layers are aligned. Denoting the lattice vectors of the pristine CrSBr unit cell as $\bm{a}$ and $\bm{b}$, the diagonal of the unit cell $\bm{u}=\bm{b}-\bm{a}$ naturally becomes one lattice vector of the moire supercell. Along the direction perpendicular to $\bm{u}$, the supercell lattice vector is $\bm{v}=m\bm{a}+n\bm{b}$, where $m$, $n$ are the smallest integers satisfying $n/m=\left(a/b\right)^2$. In CrSBr, the moire superlattice is almost incommensurate along the direction of $\bm{v}$, as with $a=3.5\ \mathrm{\AA}$ and $b=4.76\ \mathrm{\AA}$, $\left(a/b\right)^2$ is not close to the ratio of two small integers.  The qualitative difference between the $\bm{u}$ and $\bm{v}$ directions gives rise to a unique one dimensional moire pattern, illustrated in Fig.~\ref{fig:Fig5_moire_pattern}(a), for a range of angles proximate to $\theta_t$. Along $\bm{u}$ the small lattice mismatch induced by a twist $\Delta\theta\equiv\theta-\theta_t$ leads to long-wavelength moire modulation, whose period is given by 
\begin{equation}
    \lambda = \frac{ab}{\sqrt{a^2+b^2}|\sin(\Delta\theta)|}.
\end{equation}
In contrast, along $\bm{v}$ direction, the lattice is almost incommensurate even at $\theta_t$, so a small lattice mismatch does not generate a coherent moire modulation. 

The moire pattern consists of alternating bright and dark stripes, whose structures arise from the relative alignment of atomic chains along the $\bm{u}$ direction. The alignment can be visualized by examining the interface from the perspective of $\bm{u}$, as indicated by the arrow in Fig.~\ref{fig:Fig5_moire_pattern}(a). The atomic stucture of the two stripes is illustrated in Fig.~\ref{fig:Fig5_moire_pattern}(b). Within a bright stripe, the atomic chains in the top and bottom layers are aligned, corresponding to AA stacking. In contrast, in the dark stripes the chains in the two layers are shifted by half the lattice constant along the $\bm{v}$ direction, resulting in AB stacking. Fig.~\ref{fig:Fig5_moire_pattern}(c) shows that the 1D moire pattern is clearly observed in transmission electron microscopy (TEM), where the alternating AA and AB stackings are delineated by the yellow dashed lines.

\section{Discussion}
The emergence of cross-coupling between layer-localized excitons and magnons is a striking property of twin-twisted CrSBr.  In the following, we show that the $M_{\alpha,\beta} \leftrightarrow X_{\beta,\alpha}$ may arise from a previously unrecognized magnon excitation mechanism that involves interlayer charge transfer, and we explain why this phenomenon is uniquely tied to the vicinity of $\theta_t$.

\begin{figure*}[htbp]
\centering
\includegraphics[width=1.0\linewidth]{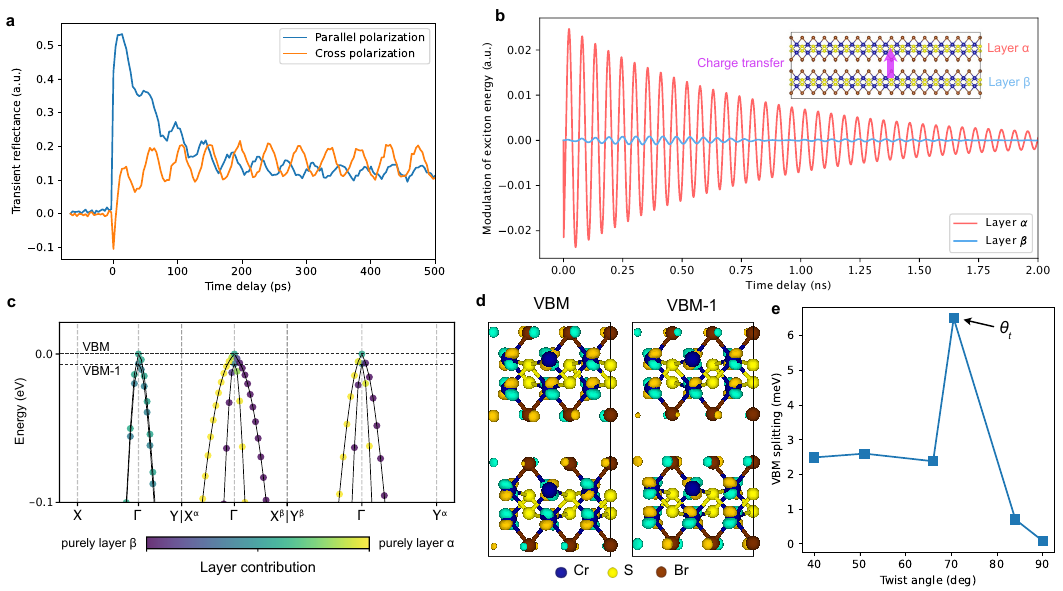}
\caption{(a) Transient reflectance of a 75$^{\circ}$ twisted sample under a 1.0 T out-of-plane field with parallel and cross polarization configuration. (b) The simulated time evolution of exciton energy modulation as a response to an impulsive spin-transfer torque associated with the charge transfer from layer $\beta$ to layer $\alpha$. (c) Calculated band structure of the twin-twisted CrSBr near VBM with spins aligned along the easy axis of each layer. In the k-path along the horizontal axis, $\Gamma$ is the Brillouin zone center; $\textrm{X}$ and $\textrm{Y}$ are the $\textrm{X}$ and $\textrm{Y}$ points of the superlattice Brillouin zone; $\Gamma$-$\textrm{X}^{\alpha,\beta}$ ($\textrm{Y}^{\alpha,\beta}$) is along  $\Gamma$-$\textrm{X}$ ($\textrm{Y}$) of the Brillouin zone for layer $\alpha$ and $\beta$ path till meeting the superlattice Brillouin zone boundary (Fig. S4). (d) DFT wavefunctions corresponding to the VBM and VBM-1 at $\Gamma$ for twin-twisted CrSBr (AA stacked). The color represents the sign of the wavefunction. (e) DFT Kohn-Sham VBM splitting ($E_{\textrm{VBM}}-E_{\textrm{VBM-1}}$ at $\Gamma$) as a function of twist angle.}
\label{fig:Fig6_theory}
\end{figure*}

We begin with compelling evidence of interlayer charge transfer. As shown in Figs.~\ref{fig:Fig3_spin_exciton_coupling}(a,b), the non-oscillatory component of the transient reflectance provides a direct probe of nonequilibrium carrier dynamics. This allows us to detect interlayer charge transfer by configuring the pump polarization along the $b$-axis of one layer (initiating nonequilibrium carrier density) and the probe polarization along the $b$-axis of the other layer (probing carrier density). Using this cross-polarized configuration, we obtain the transient reflectance signal shown as the orange curve in Fig.~\ref{fig:Fig6_theory}(a). A gradual onset in the non-oscillatory component indicates charge transfer from layer $\beta$ to layer $\alpha$. In contrast, when both pump and probe polarizations are parallel (i.e., pumping and probing the same layer), the transient reflectance exhibits a sharp change immediately after time zero, consistent with intralayer processes.

An inevitable consequence of interlayer charge transfer is the emergence of spin-transfer torque at the interface, arising from the noncollinear spin configuration between the two layers. This torque provides a magnon excitation pathway that is intrinsically interfacial in nature, as we describe below. When the pump polarization is aligned along the $b$-axis of layer $\beta$, it generates a nonequilibrium carrier population within that layer, initiating charge transfer from layer $\beta$ to layer $\alpha$. During this process, the spins in layer $\alpha$ experience a spin-transfer torque $\bm{\tau}_{\mathrm{st}} \propto \hat{\bm{S}}_{\alpha} \times (\hat{\bm{S}}_{\alpha} \times \hat{\bm{S}}_{\beta})$, where $\hat{\bm{S}}_{\alpha}$ and $\hat{\bm{S}}_{\beta}$ are the spin orientations in layers $\alpha$ and $\beta$ respectively \cite{ralphSpinTransferTorques2008a}. This torque perturbs the spin orientation in layer $\alpha$, generating magnons even though the pump polarization is nearly orthogonal to the $b$-axis of layer $\alpha$.

To further explore the role of spin-transfer torque, we simulate spin dynamics using the Landau-Lifshitz-Gilbert equation, incorporating a charge transfer process from layer $\beta$ to layer $\alpha$ under a 1.0 T out-of-plane magnetic field (see Supplementary Information Sec. 3 for details). We model the spin-transfer torque as an impulsive perturbation with a decay time of 10 ps, and compute the resulting modulation of the exciton energy using the relation $\Delta E_{\mathrm{ex}} \propto \cos\theta$, where $E_{\mathrm{ex}}$ is the exciton energy and $\theta$ is the angle between spins on the two magnetic sublattices. As shown in Fig.~\ref{fig:Fig6_theory}(b), when the pump polarization is aligned with the $b$-axis of layer $\beta$, the spin-transfer torque excites magnons in layer $\alpha$, consistent with the observed ``cross-coupling'' phenomenon $M_{\alpha,\beta} \leftrightarrow X_{\beta,\alpha}$.

Naturally, the spin-transfer-torque mechanism must coexist with the conventional bulk magnon excitation process \cite{sunDipolarSpinWave2024b,baeExcitoncoupledCoherentMagnons2022b}, which excites magnons in layer $\alpha$ ($\beta$) when the pump polarization is aligned with the $b$-axis of layer $\alpha$ ($\beta$). Cross-coupling emerges only when the spin-transfer torque mechanism dominates, which we observe exclusively near the twin twist angle $\theta_t$. 

To understand the $\theta$-dependence of cross-coupling, we use DFT to investigate the amplitude of interlayer tunneling as a function of twist angle. As shown in Fig.~\ref{fig:Fig6_theory}(c,d) the valence band maximum (VBM) splits into symmetric and antisymmetric combinations of states localized in the two layers when spins across layers cant away from antiparallel spin configurations. The difference in energy between these two states, or, VBM splitting, which serves as a measure of the interlayer hopping amplitude, is plotted as a function of twist angle \cite{shaidu2025} in Fig.~\ref{fig:Fig6_theory}(e), revealing a pronounced peak at $\theta_t$. We speculate that the enhanced splitting near $\theta_t$ occurs because at this angle the diagonal lines of the two layers align in the AA stacking configuration, producing a high density of coincidence sites. Since the spin-transfer torque is proportional to the interlayer charge transfer rate \cite{ralphSpinTransferTorques2008a}, or equivalently degree of coupling between states in the two layers, this enhanced hopping can explain why the interfacial magnon excitation mechanism dominates exclusively near $\theta_t$.

\section{Conclusion}
In summary, we have identified a special twist angle for 2D orthorhombic lattices. At this angle, the diagonal lines of the two pristine layers align, resulting in a high density of coincidence sites. For small deviations from this angle, a robust 1D moire pattern emerges. Applying this concept to the van der Waals antiferromagnet CrSBr, we demonstrate that interfacial effects become dominant due to enhanced interlayer coupling. Utilizing the strong anisotropy of CrSBr, we identify a novel interfacial magnon excitation mechanism enabled by spin-transfer torque in twin-twisted CrSBr.

Our findings open promising directions for twisted 2D magnets and, more broadly, for orthorhombic systems. In twisted 2D magnets, we highlight the significance of large twist angles in antiferromagnetically coupled spin systems. By generating noncollinear spin order, large twist angles enhance spin-dependent interlayer tunneling matrix elements that are spin-forbidden in pristine samples or at small twist angles, leading to rich phenomena in both electronic and magnetic degrees of freedom. More generally, in orthorhombic lattices, the twin-twisted angle provides a robust and tunable platform for exploring strongly correlated 1D physics. 

\section{Methods}
\subsection{Crystal growth and sample fabrication}
CrSBr crystals were prepared by a direct reaction from for elements. Chromium (99.99\%, -60 mesh, Chemsavers, USA), bromine (99.9999\%, Sigma-Aldrich, Czech Republic), and sulfur (99.9999\%, Stanford Materials, USA) were mixed in a stochiometric ratio in a quartz ampoule (35 $\times$ 220 mm) corresponding to 15 g of CrSBr. Bromine excess of 0.5 g was used to enhance vapor transport. The material was pre-reacted in an ampoule using a crucible furnace at 700 $^{\circ}$C for 12 h, while the second end of the ampoule was kept below 250 $^{\circ}$C. The heating procedure was repeated two times until the liquid bromine disappeared. The ampoule was placed in a horizontal two-zone furnace for crystal growth. First, the growth zone was heated to 900 $^{\circ}$C, while the source zone was heated at 700 $^{\circ}$C for 25 h. For the growth, the thermal gradient was reversed and the source zone was heated from 900 to 940 $^{\circ}$C and the growth zone from 850 to 800 $^{\circ}$C over a period of 7 days. The crystals with dimensions up to 5 $\times$ 20 mm were removed from the ampule in an Ar glovebox.

Few-layer CrSBr flakes were mechanically exfoliated from bulk crystals onto 300 nm SiO$_2$/Si substrates under inert conditions in a nitrogen-filled glovebox. The layer number was first identified by the optical contrast and further confirmed by atomic force microscopy (AFM). The twisted CrSBr structure was assembled via the polymer-assisted dry transfer method. The twist angle was determined from the long edges of the top and bottom flakes and precisely controlled by a motorized rotation stage. Samples for transmission electron microscopy (TEM) were transferred to TEM grids with silicon nitride membranes.

\subsection{Transient reflectance microscope}
The transient reflectance experiments were carried out with 710-nm pump and wavelength-tunable probe pulses generated from the ORPHEUS-TWINS optical parametric amplifiers pumped by the Light Conversion CARBIDE Yb-KGW laser amplifier operating at the repetition rate of 300 kHz. With a 50x objective lens (N.A. = 0.50), both beams were focused onto the sample surface with approximate spot sizes of 2 $\mu$m unless otherwise stated, with incident laser powers fixed at $\approx$ 1 $\mu$W. The pump laser pulses were modulated at 20 kHz with a chopper and the transient reflectance signals were measured with a lock-in amplifier (MFLI, Zurich Instruments). The external magnetic field was applied by the superconducting coil inside the Quantum Design OptiCool cryostat.

\subsection{Photoluminescence spectroscopy}
The photoluminescence spectroscopy were carried out with a 633-nm He-Ne laser. The laser beam was focused onto the sample surface with a 50x objective lens. The reflected beam and photoluminescence were collected by the same ojective lens, directed into the Horiba iHR-320 spectrometer and measured by the Teledyne Princeton Instruments PyLoN CCD camera. A longpass color filter was used to block the 633-nm beam from the spectrometer.

\subsection{First-principles calculations}
We use Quantum Espresso \cite{giannozzi2009a,giannozzi2017a}, optimized norm-conserving pseudopotentials (ONCVPSP v0.4) from PseudoDojo \cite{hamann2013,vansetten2018a} and a wavefunction plane wave kinetic energy cutoff of 94 Ry for all calculations. Total energies are converged to within 1$\times 10^{-8}$ Ry/atom, and all Hellmann-Feynman forces are converged to be below 3$\times 10^{-2}$ eV/$\text{\AA}$ on each atom. We use density functional theory (DFT) within the generalized gradient approximation of Perdew, Burke, and Ernzerhof \cite{perdew1996b}, including Grimme's DFT-D3 van der Waals with BJ damping \cite{grimme2010a} for all calculations. We note that the exchange-correlation functionals used can influence the predicted eigenstate wavefunction and band structure of CrSBr. However, for AA stacked structure, the VBM energy splitting only increases by 1 meV when using Heyd–Scuseria–Ernzerhof screened hybrid functional \cite{heyd2003}, suggesting PBE is sufficient for this work.

We use BerkeleyGW \cite{deslippe2012} to perform \textit{GW}-BSE calculations using a PBE starting point with scalar ferromagnetic configurations. We compute the inverse dielectric matrix with the frequency dependence modeled using the Godby-Needs generalized plasmon pole model \cite{deslippe2013,godby1989}. The inverse dielectric matrix $\varepsilon^{-1}_{\textbf{GG'}}(\textbf{q},w)$ is constructed using 1023 bands and 10 Ry cutoff. We use a slab Coulomb truncation to reduce interactions between periodic images along the z direction \cite{ismail-beigi2006}. In our dielectric function calculation, we apply a scissor shift of 0.686 eV for conduction bands and -1.233 eV for valence bands, following values are based on a $G_{0}W_{0}$@PBE calculation on regular FM bilayer CrSBr. The dielectric function is sampled on a uniform $\Gamma$-centered $8\times6\times1$ k-mesh. The BSE matrix elements are computed on a uniform $\Gamma$-centered $8\times6\times1$ coarse k-grid with 36 valence bands and 24 conduction bands, then interpolated onto a rectangular patch of $40\times30\times1$ centered around $\Gamma$ point \cite{alvertis2023d} with 6 valence bands and 6 conduction bands. We use the momentum operator to construct all dipole transition matrix elements.

\section{Acknowledgements} 
We acknowledge support of the Quantum Materials program under the Director, Office of Science, Office of Basic Energy Sciences, Materials Sciences and Engineering Division, of the U.S. Department of Energy, Contract No. DE-AC02-05CH11231. J.O and Y.S received support from the Gordon and Betty Moore Foundation's EPiQS Initiative through Grant GBMF4537 to J.O. at UC Berkeley. F.M. and J.Y. acknowledge support from the U.S. Department of Energy, Office of Science, Office of Basic Energy Sciences, Materials Sciences and Engineering Division under contract DE-AC02-05-CH11231 (Organic-Inorganic Nanocomposites KC3104). S.K. and J.B.N. acknowledge support from the Theory of Materials FWP supported by the U.S. Department of Energy, Office of Science, Office of Basic Energy Sciences, under Award No. DE-SC0020129. S.K. and J.B.N. also acknowledge computational resources provided by the National Energy Research Scientific Computing Center (NERSC), supported by the Office of Science of the Department of Energy operated under Contract No. DE-AC02-05CH11231 using NERSC Award No. ERCAP0033609. K.X. and A.M. acknowledge cryo-EM support from the US Department
of Energy, Office of Basic Energy Sciences, Division of Materials Science and Engineering under
contract DE-AC02-76SF00515. K.X. and A.M. also acknowledge the
Stanford Nano Shared Facilities and the Stanford Nanofabrication Facility. Z.S. was supported by the ERC-CZ programme (project no. LL2101) from the Ministry of Education Youth and Sports and used large infrastructure from project reg. no. CZ.02.1.01/0.0/0.0/15\_003/0000444 financed by the European Regional Development Fund.

\section{Author contributions}
Y.S. and J.O. designed research. Y.S. carried out optical measurements under the supervision of J.O. Bulk crystals were synthesized and characterized by A.S. under the supervision of Z.S. F.M. fabricated all twisted samples under the supervision of J.Y. K.X. and H.Z. performed transmission electron microscopy under the supervision of A.M. and R.R. S.K. performed first-principle calculations under the supervision of J.B.N. Theoretical analysis was performed by Y.S., S.K., J.B.N. and J.O. Y.S., S.K., J.B.N. and J.O. wrote the paper.

\section{Competing interests}
The authors declare no competing interests.

\section{Data Availability}
All data sets supporting the conclusions of the paper and Supplementary Information are shared in a public accessible repository.

\bibstyle{apsrev4-1}

\bibliography{Main_Text_References}

\end{document}


\title{Electron-magnon coupling at the interface of a ``twin-twisted'' antiferromagnet: supplementary information}
\maketitle

\section{Photoluminescence from small-twist-angle samples}
\begin{figure}[H]
    \centering
    \includegraphics[width=0.8\linewidth]{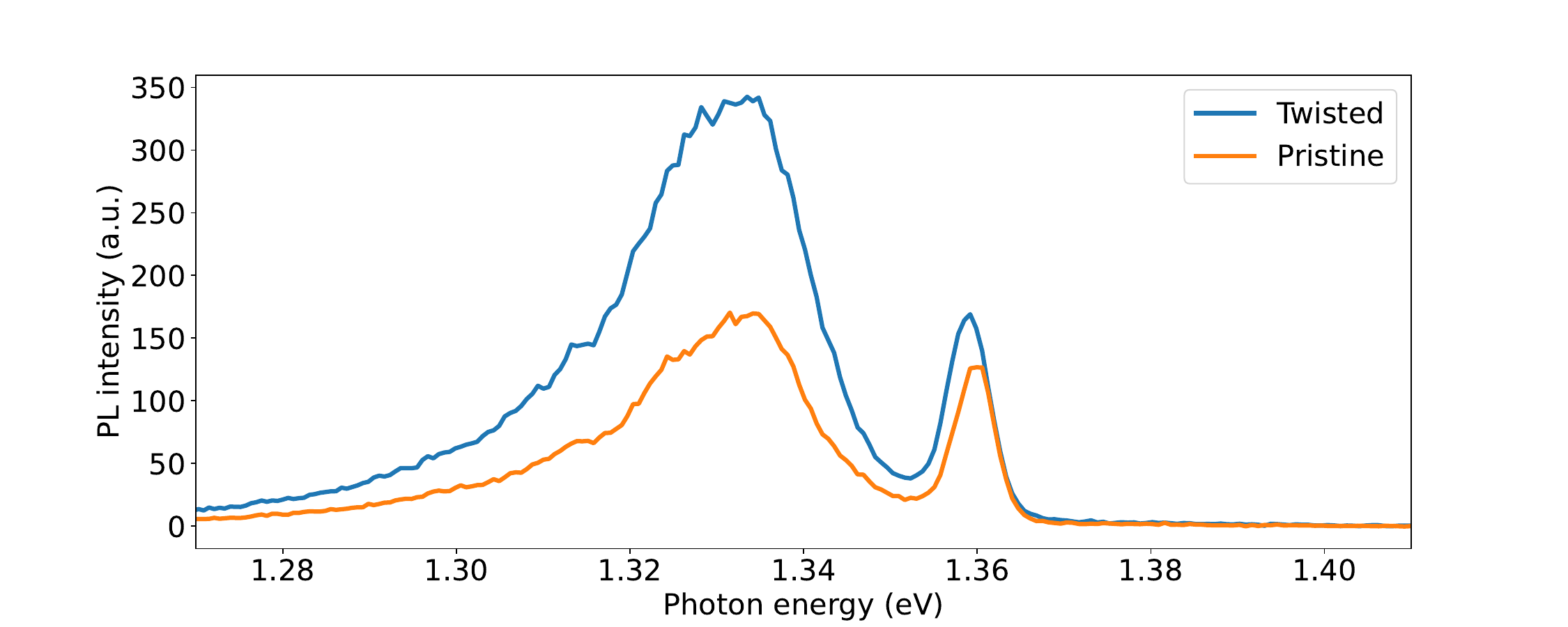}
    \caption{The comparison of PL spectra between overlapped regions (twisted) and individual layers (pristine) for 8$^{\circ}$-twisted CrSBr.}
    \label{fig:PL_8deg}
\end{figure}

In small-twist-angle samples, we do not observe the splitting of PL peaks. In Fig.~\ref{fig:PL_8deg}, we use the PL data from 8$^{\circ}$-twisted CrSBr as an example.

\section{Background subtraction}
\begin{figure}[H]
    \centering
    \includegraphics[width=0.8\linewidth]{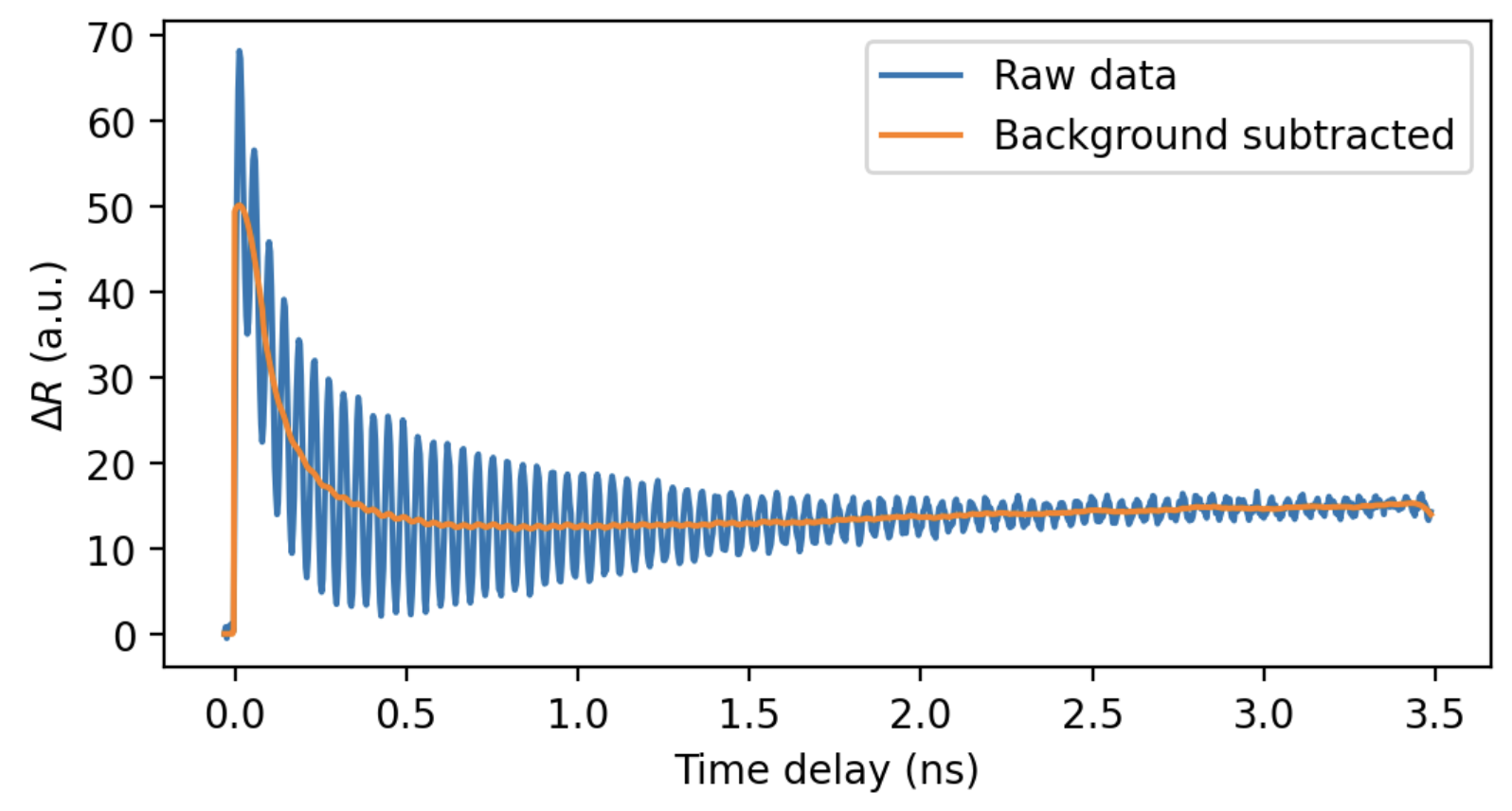}
    \caption{Comparison between the raw data and subtracted background in the transient reflectance data processing.}
    \label{fig:background_subtraction}
\end{figure}

In the data processing for the spin wave spectroscopy, we subtracted the non-oscillatory component of the transient reflectance signal using the Savitzky-Golay filter. Fig.~\ref{fig:background_subtraction} shows the background subtracted from the raw data.

\section{Spin dynamics simulations}
We use the Landau-Lifshitz Gilbert equation below to simulate the spin dynamics induced by the spin-transfer torque. 
\begin{equation}
    \frac{\partial \bm{M}}{\partial t} = -\gamma \bm{M}\times\bm{H}_{\textrm{eff}}-\alpha_0 \bm{M}\times(\bm{M}\times\bm{H}_{\textrm{eff}})+\tau_{st},
\end{equation}
where $\gamma$ is the gyromagnetic ratio, $\alpha_0$ is the damping constant and $\tau_{st} = \lambda \bm{M}_{\alpha}\times(\bm{M}_{\alpha}\times\bm{M}_{\beta})$ is the spin-transfer torque. In the simulation, we assume both layer $\alpha$ and $\beta$ are bilayer CrSBr. The twist angle between the two bilayers is 75$^{\circ}$. To be consistent with the observation that magnons are localized in each layer, we let layer $\alpha$ and $\beta$ have slightly different amplitudes of the in-plane easy-axis anisotropy with $K_{x,\alpha} = 2.00\times10^4$ J/m$^3$ and $K_{x,\beta} = 1.80\times10^4$ J/m$^3$. The hard-axis anisotropy along the $c$-axis is $K_z = 1.10\times10^5$ J/m$^3$ and the antiferromagnetic exchange interaction inside each bilayer is $J = 6.48\times10^4$ J/m$^3$ for both layer $\alpha$ and $\beta$. The interfacial interaction between layer $\alpha$ and $\beta$ is $J_{int} = -6.48\times10^2$ J/m$^3$ and the saturation magnetization is $M_s = 2.05\times10^5$ A/m. We consider a charge transfer from layer $\beta$ to layer $\alpha$, so $\tau_{st}$ is applied on layer $\alpha$. $\tau_{st}$ is modeled as an impulsive perturbation with a decay time of 10 ps.

\section{Ab inito calculations of twin-twisted C\MakeLowercase{r}SB\MakeLowercase{r}}
\begin{figure}
    \centering
    \includegraphics[width=0.6\linewidth]{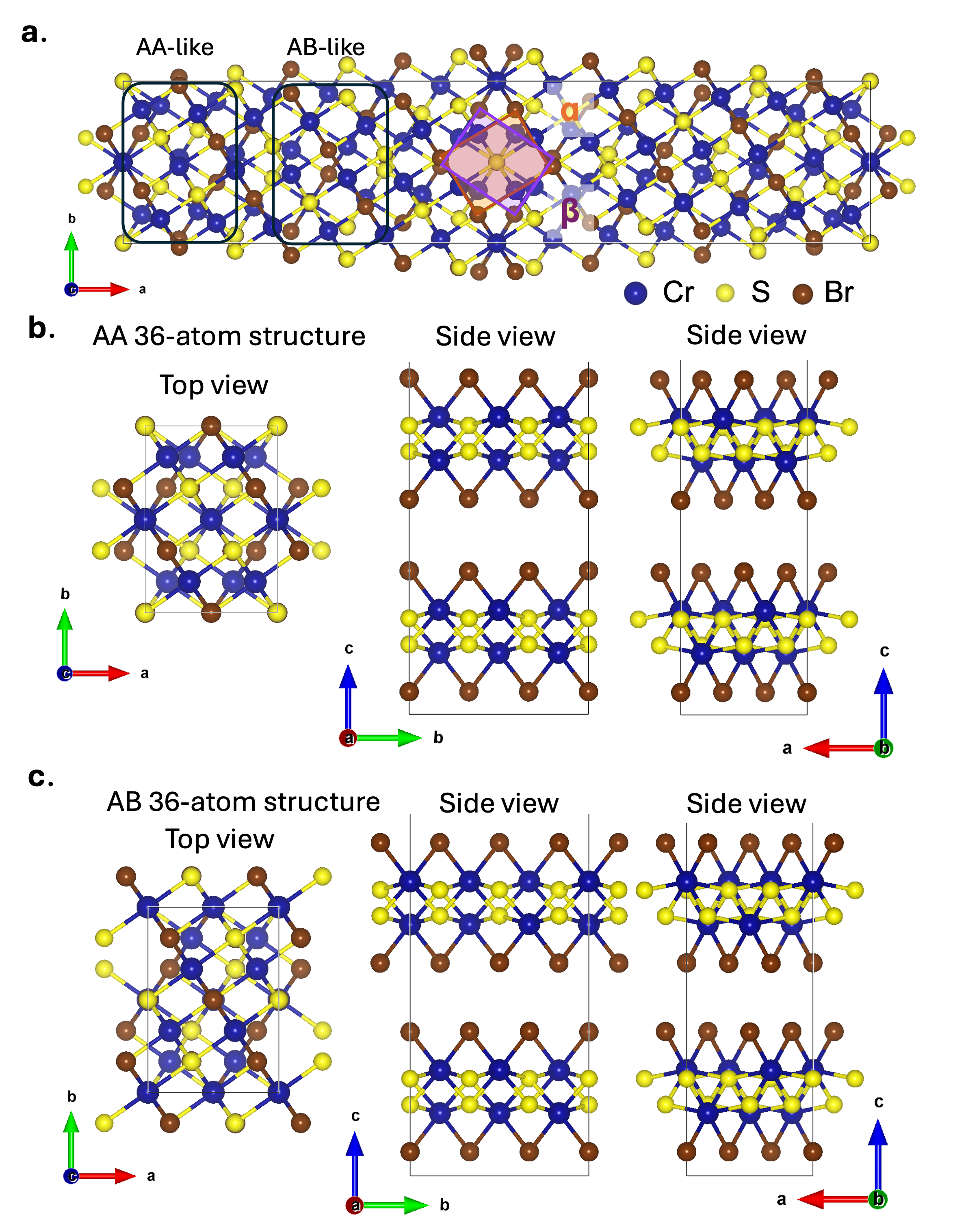}
    \caption{\textbf{(a)} Top view of a relaxed 240-atom supercell that models experimental samples near $\theta_{c}$. There are two regions: an AA-like region and an AB-like region. The transparent pink rhombus outlines the moiré unit cell. The two shaded rectangles in the center highlight the two primitive cells from $\alpha$ (orange; upper) and $\beta$ (purple; lower) CrSBr layers. 36-atom \textbf{(b)} AA and\textbf{ (c)} AB stacking in top view and side views along the \textit{a}- and \textit{b-directions.}}
    \label{fig:local-stacking}
\end{figure}

To understand the electronic structure of twin-twisted CrSBr, we use density functional theory (DFT) to obtain relaxed structures, band structures, and wavefunctions,  and the \textit{GW}-Bethe-Salpeter equation (\textit{GW}-BSE) calculations to capture the first few lowest bright excitations including electron-hole interactions for local bilayer stacking configurations in the Moiré pattern associated with $\theta_{c}=2 \tan(\frac{b}{a})$. The resulting Moiré pattern near angles of $\theta_{c}$  has two near-commensurate regions (Fig. \ref{fig:local-stacking} \textbf{a}), which we model separately as AA and AB stacked structures (Fig. \ref{fig:local-stacking} \textbf{b,c}). Viewed along the superlattice \textbf{\^a} axis, we define the stacking based on the relative alignment of nearest-neighbor Br atoms across the layers: in the AA configuration, they are perfectly aligned along \textbf{\^b} direction; in the AB configuration, they are displaced by a fractional vector along \textbf{\^b} direction. We focus on results from the AA local stacking configuration (Fig. \ref{fig:local-stacking} \textbf{b}), as other configurations (for example, AB shown in Fig. \ref{fig:local-stacking} \textbf{c}) yield comparable band structures and excitation characteristics (shown in Fig. \ref{fig:band-diff}). The details of supercell constructions can be found in the following.
\subsection{Supercell constructions}
A standard bilayer CrSBr consists of two perfectly aligned monolayers, stacked with zero in-plane displacement, with lattice parameters $a = $ 3.508 $\text{\AA}$ and $b = $ 4.763 $\text{\AA}$. To construct the local configuration AA and AB, we first build the lower monolayer by making a supercell with one lattice vector being the diagonal of the primitive unit cell ($\boldsymbol{u^{lower}}=\boldsymbol{a}-\boldsymbol{b}$) and another lattice vector being $\boldsymbol{v^{lower}}=2\boldsymbol{a}+\boldsymbol{b}$; we further construct the upper layer by making a supercell with two lattice vectors being $\boldsymbol{u^{upper}}=\boldsymbol{a}-\boldsymbol{b}$ and $\boldsymbol{v^{upper}}=-2\boldsymbol{a}-\boldsymbol{b}$. Here we strain the lattice parameters of the monolayer primitive CrSBr cell, $a = $3.426 $\text{\AA}$ (2.3\% contraction), $b = $4.846 $\text{\AA}$ (1.7\% expansion), such that the supercell is orthorhombic. The AA structure is constructed by stacking these two monolayers together, and the AB structure is built by stacking one lower layer supercell and one upper supercell that is shifted by half $\boldsymbol{v^{upper}}$. The vacuum distance is 15 $\text{\AA}$. After construction, the supercells are relaxed in DFT, which details is described in Methods section.

\subsection{Spin order and electronic structures}
The tunneling resistance measurements \cite{chenTwistassistedAllantiferromagneticTunnel2024} indicate a non-antiparallel magnetic configuration for twin-twisted CrSBr, which suggests that interlayer hybridization is not spin-forbidden as in standard AFM CrSBr bilayer for $\theta=0$. We assume the magnetic anisotropy dominates in each CrSBr monolayers, therefore, we use non-collinear spin configurations in DFT calculations, where spins of each layer are aligned along their magnetic easy axis and the spin angle between two layers is roughly $\theta_{c}$. Changing the spin angle barely changes the character of electronic structures and wavefunctions, it only affects the degree of interlayer hybridization, which is reflected in the energy splitting values. The DFT-PBE VBM splitting can be roughly approximated as $E_{VBM}-E_{VBM-1} \propto \cos(\theta_{spin}/2)$. To save computational cost, we use a scalar spin ferromagnetic configuration in the \textit{GW}-BSE calculations. 

\begin{figure}
    \centering
    \includegraphics[width=0.8\linewidth]{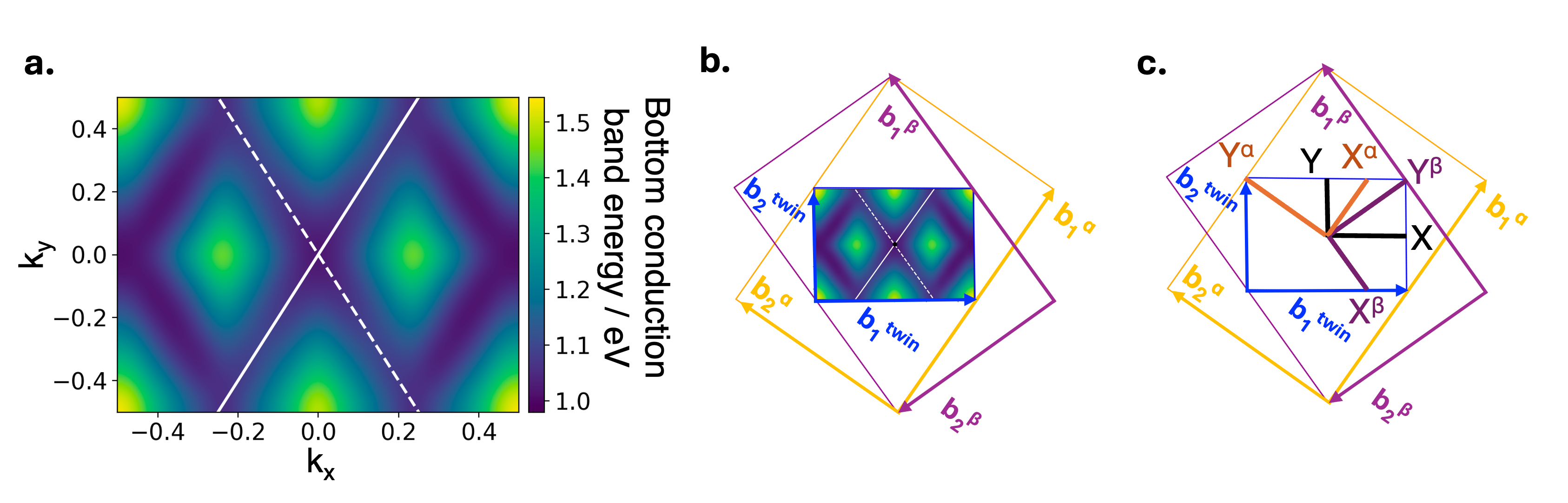}
    \caption{\textbf{(a) }DFT-PBE lowest conduction band energy map in AA stacking with scalar ferromagnetic configuration; the white line shows $k_{y}=2k_{x}$ and white dashed line shows $k_{y}=-2k_{x}$, \textbf{(b)} Brillouin zone of AA (blue) and primitive cell of each layer ($\alpha$ layer in yellow and $\beta$ layer in purple). (c) Brillouin zone with high-symmetry points and paths.}
    \label{fig:anisotropy}
\end{figure}

\begin{figure}
    \centering
    \includegraphics[width=0.7\linewidth]{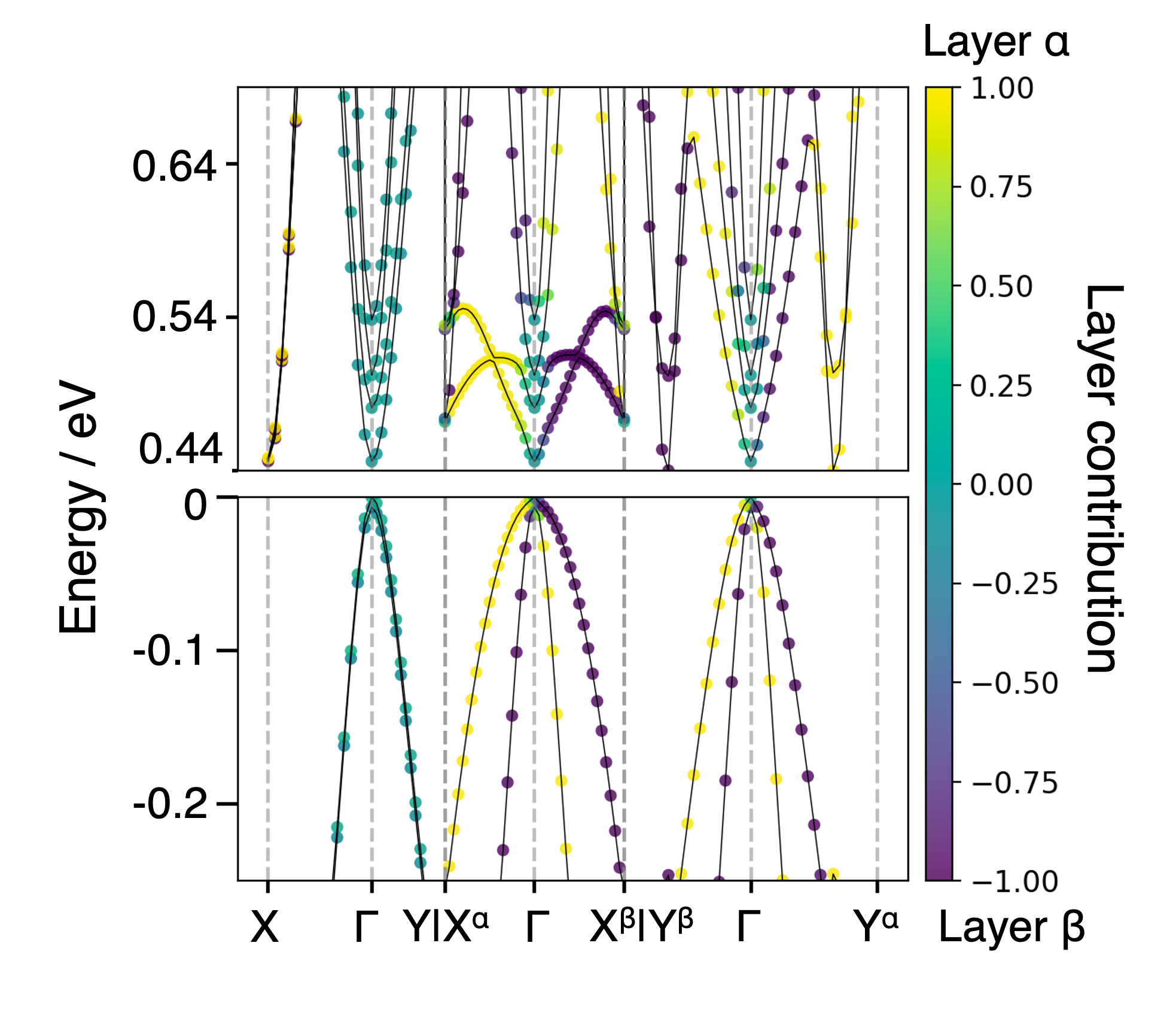}
    \caption{Layer-projected DFT-PBE band structure in AA stacking structure with noncollinear spin configuration where spins of each layer are aligned along its magnetic easy axis and the angle between spins across layers is 70.5 \degree. We set reference zero energy as VBM eigenvalue.}
    \label{fig:proj}
\end{figure}

\begin{figure}
    \centering
    \includegraphics[width=0.8\linewidth]{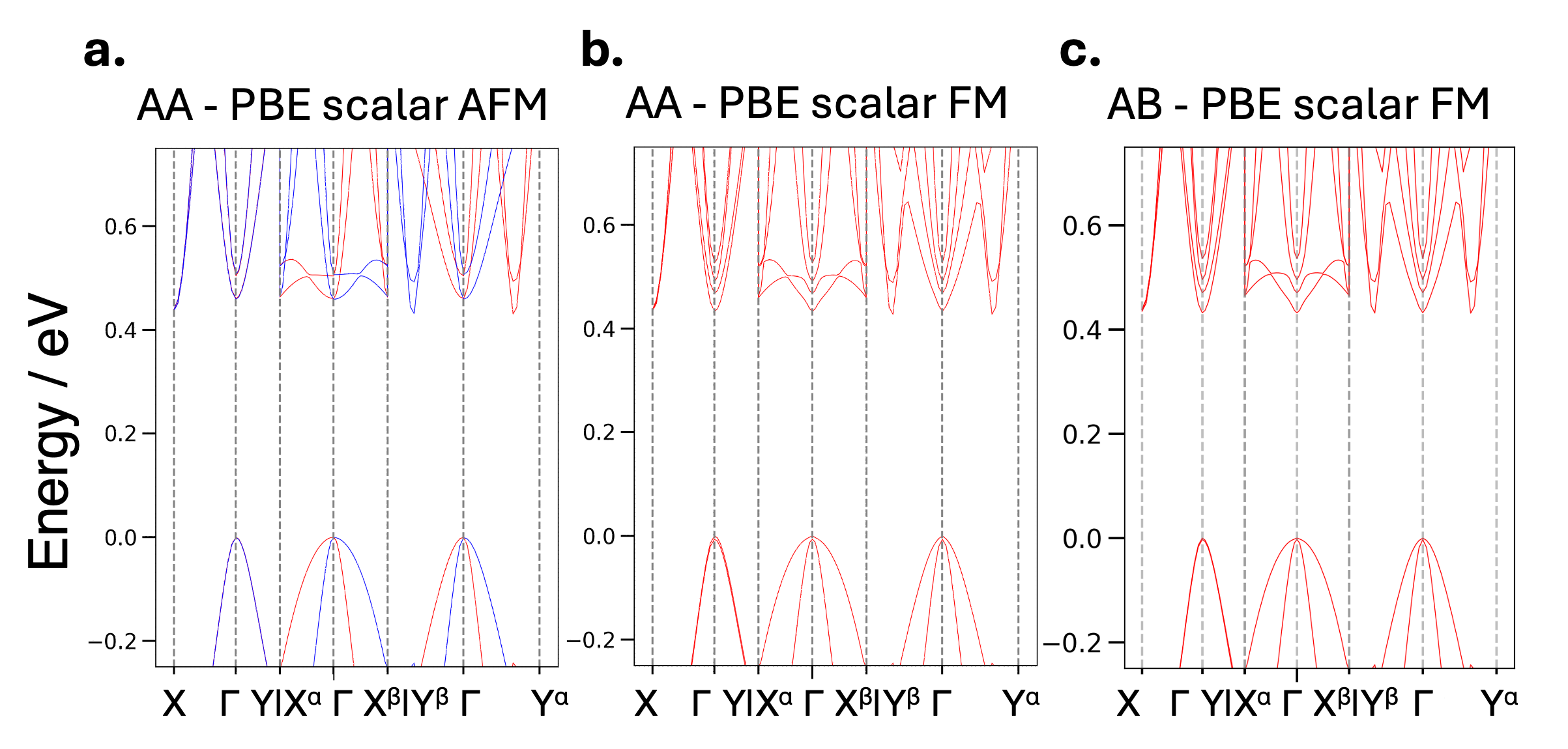}
    \caption{DFT-PBE calculated band structure in AA stacking with scalar \textbf{(a)} antiferromagnetic and \textbf{(b)} ferromagnetic configuration and in AB stacking with scalar \textbf{(c)} ferromagnetic configuration. The band structure is very similar between \textbf{(b)} and \textbf{(c)}.}
    \label{fig:band-diff}
\end{figure}

Fig. \ref{fig:proj} shows the DFT band structure of the local stacking AA structure, with the color representing layer contribution. The lowest one-particle interband transition appears at the $\Gamma$ point in the reciprocal space, with contributions from both layers, similar to standard CrSBr bilayers ($\theta=0$). The band structure also exhibits an anisotropic lowest conduction band (shown in Fig. \ref{fig:anisotropy}), arising from band folding from a primitive cell of standard CrSBr, where the lowest conduction band is nearly flat along $\Gamma$-$X$ and dispersive along $\Gamma$-$Y$ in the primitive Brillouin zone. Therefore, states on the original $\Gamma$-$X$ line for upper CrSBr primitive cell are folded into $k^{AA}_{y} = 2 k^{AA}_{x}$ for $|k^{AA}_{x}| < 0.25$, $k^{AA}_{y} = 2 (k^{AA}_{x}-0.25)-0.5$ for $k^{AA}_{x} > 0.25$, and $k^{AA}_{y} = 2 (k^{AA}_{x}+0.5)$ for $k^{AA}_{x} <- 0.25$ in the supercell Brillouin zone. Along these paths, the lowest conduction band is nearly flat and mostly dominated by the upper layer as shown in Fig. \ref{fig:anisotropy}. Similarly, the lowest conduction states along paths $k^{AA}_{y} = -2 k^{AA}_{x}$ for $|k^{AA}_{x}| < 0.25$, $k^{AA}_{y} = -2 (k^{AA}_{x}-0.25)+0.5$ for $k^{AA}_{x} > 0.25$, and $k^{AA}_{y} = -2 (k^{AA}_{x}+0.5)$ for $k^{AA}_{x} <- 0.25$ have lower energy and are localized in the lower layer. From Fig. \ref{fig:proj}, we find that interlayer hybridization near band edges only becomes noticeable when near $k_{x}=0$ or $k_{y}=0$. \par
\subsection{Excited state properties of AA stacked structure }
\begin{figure}
    \centering
    \includegraphics[width=0.9\linewidth]{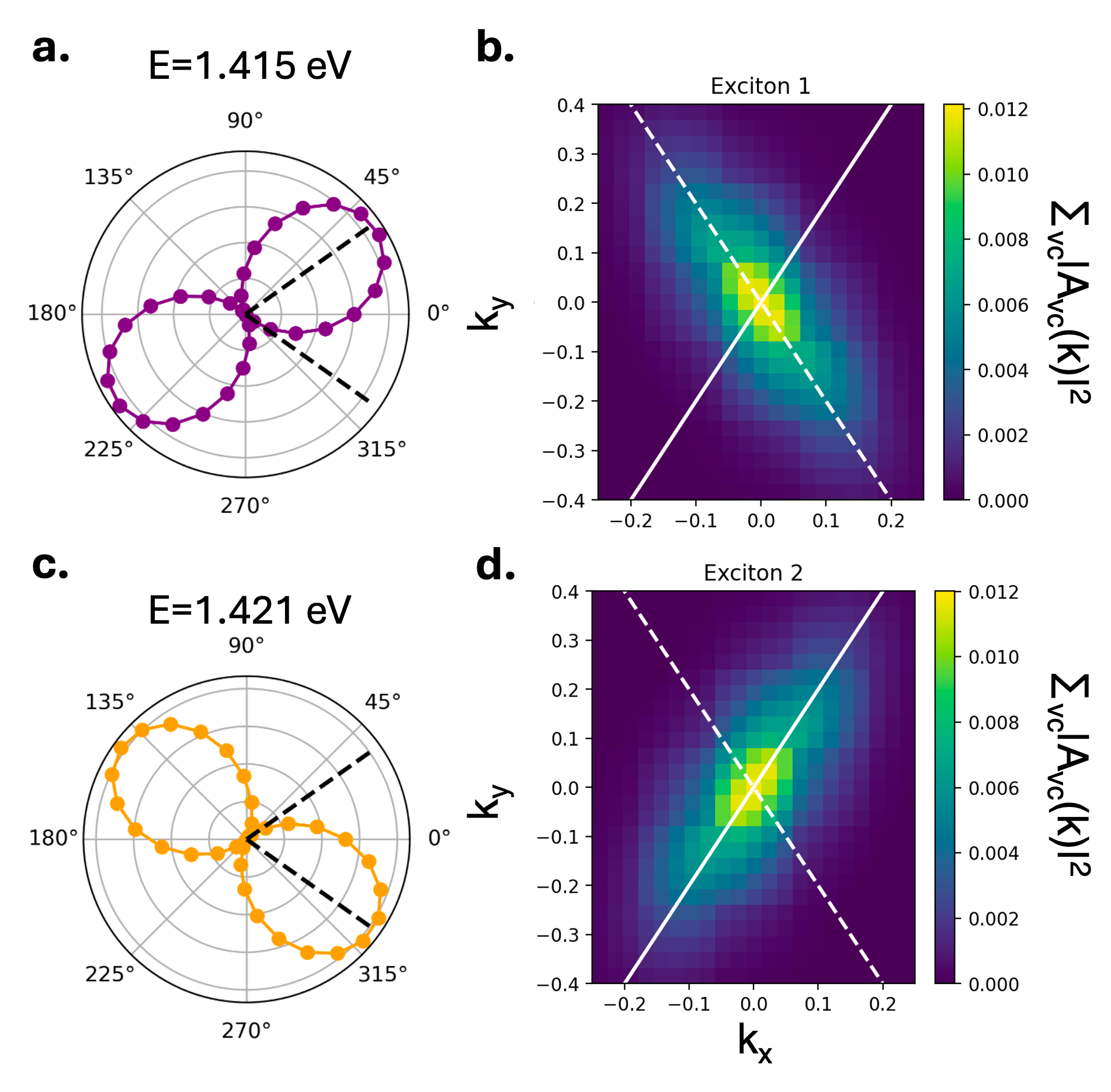}
    \caption{The polarization dependence of the lowest (\textbf{a}) exciton and the second (\textbf{c}) lowest exciton in energy. The black dashed line highlights the easy axis of each layer. Exciton coefficient $\sum_{v,c}|A^{S}_{vck}|^{2}$ for the first lowest exciton (\textbf{b}) and the second lowest exciton (\textbf{d}) in the reciprocal space. The white line shows $k_{y}=2k_{x}$ and white dashed line shows $k_{y}=-2k_{x}$.}
    \label{fig:pl}
\end{figure}

Fig \ref{fig:pl} \textbf{(a,c)} from our \textit{GW}-BSE calculations reveal that the lowest two excitons are bright and exhibit polarizations along the easy axis of the upper and lower layers, respectively, in good agreement with our experimental findings. The polarizations of these states originates from electron-hole interaction, which mixes vertical one-particle transitions where only one layer dominates in regions of reciprocal space near the band edges. We plot the sum of coefficients of $\sum_{v,c}|A^{S}_{vck}|^{2}$ that describe the sum of contributions of transitions between occupied valence state $v$ and unoccupied conduction state $c$ at wavevector $k$ to excited state $S$. The exciton coefficient $A^{S}_{vck}$ for optical transitions is defined as,
\begin{equation}
    \Psi^{S} (r_{e},r_{h}) = \sum_{vck} A^{S}_{vck}\psi_{ck}(r_{e})\psi^{*}_{vk}(r_{h})
\end{equation}
Fig. \ref{fig:pl} \textbf{b} and \textbf{d} shows $\sum_{v,c}|A^{S}_{vck}|^{2}$  in the reciprocal space respectively for the first exciton ($S=1$) and the second exciton (S=2). Most contributions to the first exciton (or the second exciton) come from particular reciprocal regions centered along $k^{AA}_{y} = -2 k^{AA}_{x}$ for $|k^{AA}_{x}| < 0.25$ (or $k^{AA}_{y} = 2 k^{AA}_{x}$ for $|k^{AA}_{x}| < 0.25$), where transitions that occur between states are dominated by the upper layer $\beta$ (or the lower layer $\alpha$). In a single-particle picture, the polarization of the lowest two excitations would be dominated only by the wave functions only at $\Gamma$ and would be each along one of the two angle bisectors for the easy axes of the two layers. We hypothesize that an exciton energy splitting of the lowest two excitons, as observed in Fig. 1a of the main text, may originate from the differences in the atomic structure of the upper and lower layers and associated asymmetric interlayer hybridization between layers, which will be an interesting subject of future study. 

\section{Moire pattern associated with other angles}
We use the twist\_layers code \cite{shaidu2025} to generate moire pattern supercells with other twist angles (40 \degree, 51 \degree,  66 \degree, 84 \degree, and 90 \degree) and relax them in DFT. We choose these angles because their moire patterns well represent possible patterns in the larger twist angle regime, and the total number of atoms in these supercells is relative small, saving computationally cost. The valence band maximum at $\Gamma$ point of twisted CrSBr splits into two states when spins are not aligned antiparallel, and this energy difference is used as a measure of interlayer hybridization. Here we show the DFT-PBE-calculated VBM energy splitting in collinear ferromagnetic configurations in SI Table \ref{tab:vbs_angle}. 

\begin{table}[htbp]
    \centering
    \begin{tabular}{|c|c|c|c|c|c|c|}
        \hline
         Angle/\degree&  40&  51&  66&  70.5&  84& 90\\\hline
 \# of atoms& 96& 132& 156& 36& 132&144\\ \hline 
         VBM energy splitting / meV&  2.5&  2.6&  2.4&  6.5&  0.7& 0\\ \hline
    \end{tabular}
    \caption{DFT-PBE calculated VBM energy splitting with respect to twist angles.}
    \label{tab:vbs_angle}
\end{table}

\begin{figure}[H]
    \centering
\includegraphics[width=0.65\linewidth]{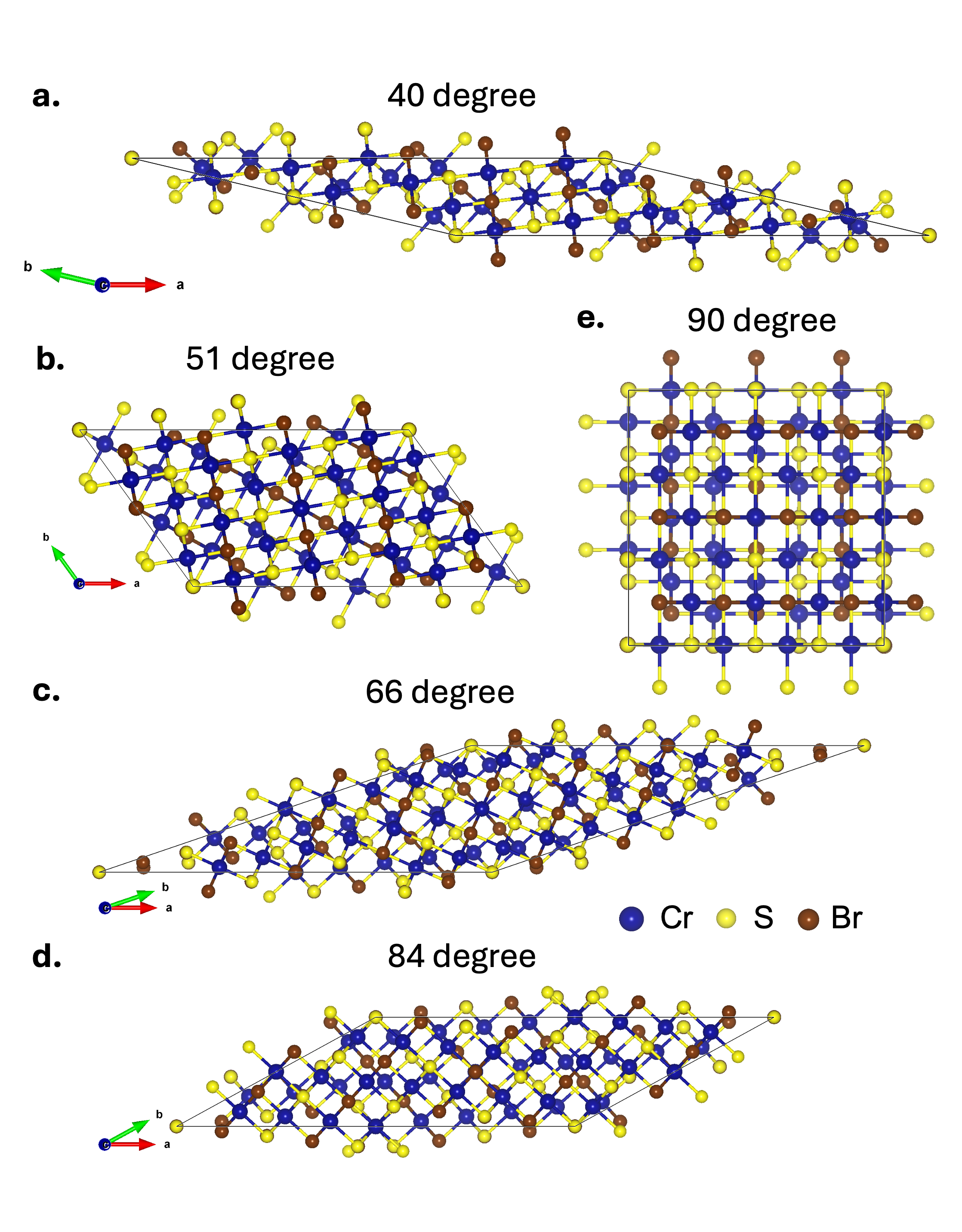}
    \caption{Top views of moire pattern supercells of twist angles, \textbf{(a)}40 \degree, \textbf{(b)} 51 \degree, \textbf{(c)} 66 \degree,\textbf{ (d)} 84 \degree, and \textbf{(e)} 90 \degree}
    \label{fig:angle}
\end{figure}
\bibstyle{apsrev4-2}

\bibliography{SI_text_citation}